\documentclass[epj]{svjour}
\usepackage{graphics}

\usepackage{latexsym}
\usepackage[ansinew]{inputenc}
\usepackage{amsmath}
\usepackage{textcomp}
\usepackage{amsfonts}
\usepackage{amssymb}
\usepackage{mathrsfs}
\usepackage{color} 
\usepackage{graphicx}
\usepackage{subfigure}
\usepackage{bm}

\begin{document}
\title{Networks of noisy oscillators with correlated degree and frequency dispersion}
\author{B. Sonnenschein\inst{1,2,}\thanks{sonne@physik.hu-berlin.de} \and F. Sagu\'es\inst{3} \and L. Schimansky-Geier\inst{1,2}\\
}                     
%
%
\institute{
  \inst{1} Department of Physics, Humboldt-Universit\"at zu Berlin, Newtonstrasse 15, 12489 Berlin, Germany\\
  \inst{2} Bernstein Center for Computational Neuroscience Berlin, Philippstrasse 13, 10115 Berlin, Germany\\
  \inst{3} Departament de Qu\'imica F\'isica, Universitat de
  Barcelona, Mart\'i i Franqu\`es 1, 08028 Barcelona, Spain}
%
%
\abstract{
We investigate how correlations between the diversity
    of the connectivity of networks and the dynamics at their nodes
  affect the macroscopic behavior. In particular, we study the
  synchronization transition of coupled stochastic phase oscillators
  that represent the node dynamics. Crucially in our work, the
  variability in the number of connections of the nodes is correlated
  with the width of the frequency distribution of the oscillators. By
  numerical simulations on Erd\"os-R\'enyi networks, where the
  frequencies of the oscillators are Gaussian distributed, we
  make the counterintuitive observation that an increase in the
  strength of the correlation is accompanied by an increase in the
  critical coupling strength for the onset of synchronization. We
  further observe that the critical coupling can solely depend on the
  average number of connections or even completely lose its dependence
  on the network connectivity. Only beyond this state, a weighted
  mean-field approximation breaks down. If noise is present, the
  correlations have to be stronger to yield similar
    observations.
} 
\titlerunning{Correlated degree and frequency dispersion}
\authorrunning{B. Sonnenschein \emph{et al.}}
\maketitle
\section{Introduction}
\label{intro}
In the last decade network science has become a field of research with
increasing importance. This is mainly due to the fact that in
principle any kind of coupling structure can be mapped to {a network
  of specific complexity}.  {In this way, one aims for
  understanding fundamental properties that networks with given
  structure may have in common}. Besides  {a} large variety of
locally connected networks, the two most prominent examples were
coined by Watts and Strogatz \cite{WaStr98}, and Barab\'asi and Albert
\cite{BarAl99}, who showed that various networks can be divided into
so-called small-world or scale-free
networks, respectively. In small-world networks, all the nodes
typically have the same range of neighbor connections plus a
few random shortcuts are established.
In contrast, scale-free networks are characterized by a significant
amount of ``hub'' nodes with a very large number of connections.
Network science further owes its popularity to the growing relevance
of interdisciplinary topics and to the advances in computer technology
and science \cite{KlEguToSan03}.

One important topic is the study of the interplay between network
topology and dynamics on the nodes, as impressively reviewed in
\cite{AlBar02,New03,BocLaMoChHw06,BaBartVesp08,ArDiazKuMoZh08}.  We
pose the question, how correlations between connectivity and dynamics on the
microscopic level affect the macroscopic behavior of a network. Such
a connection between the numbers of links in a network and the functional
ability of the node dynamics is evident and possibly caused by
various reasons, for example by limited energy supply, restricted
space or chemical resources, etc. Indeed, various types of neurons 
differ in the typical number of connections and firing rates \cite{VarChPanHaChk11}.

In order to approach the problem, we investigate the synchronization
transition of phase oscillators in complex networks. The phenomenon of
synchronization suits a \\*benchmark by virtue of its importance as a
paradigmatic emergence of collective behavior, as outlined for
instance in \cite{PikRosKu03,BalJaPoSos10}.

Only recently, G\'omez-Garde\~nes \emph{et al.} showed that a special type of
such a correlation can lead to an onset of synchronization resembling
a first-order phase synchronization \cite{GogaGoArMo11}. This is
remarkable, because the synchronization transition was always found to
be of second order, if one only considers how different network
topologies affect the dynamics. They instead identified the natural
frequency of each node with its individual degree $\omega=k$, i.e.,
its number of connections. Furthermore, they interpolated between
Erd\"os-R\'enyi random networks and scale-free networks. In this way
it was found that the first-order nature of the synchronization
transition appears only in scale-free networks; those networks are
characterized by an unlimited dispersion of degrees. Hence, it was
shown that a positive correlation between the dynamics of the
oscillators and the large heterogeneity of the network has a drastic
effect on the onset of synchronization.

In this paper, we consider a more general correlation between
the degree and the frequency distributions; we relate the diversity of the
frequencies to the degrees. Two different
settings are generated, namely either positive or negative correlations between
the degree of a node and how broad its oscillatory frequency varies from the
mean one. We explore whether this kind of correlation with fixed average
natural frequency is enough to yield a notable impact on the
synchronization transition. In particular, we focus on the
Erd\"os-R\'enyi random network model, which often serves as an important
  benchmark \cite{VarChPanHaChk11}. We show by simulations that the correlations can either
support or impede the synchronizability. The results are supported by help of a weighted mean
  field theory \cite{SonnSchi12} which allows to formulate quantitative dependencies for
  the critical coupling within the validity of the
  approximative theory. We conclude by a qualitative discussion of our
  findings.

\section{How degree-frequency correlations affect the synchronization
  transition}
The most prominent model in studying synchronization phenomena is the
Kuramoto model \cite{Kur84}
\begin{equation}
  \dot{\phi}_i(t)=\omega_i+\frac{\kappa}{N^q}\sum_{j=1}^{N}A_{ij}\sin\left(\phi_j(t)-\phi_i(t)\right)+\xi_i(t),
\label{ourmodel}
\end{equation}
where $i=1,\ldots,N$, with natural frequency $\omega_i$ and phase
$\phi_i(t)$ of oscillator $i$ at time $t$, respectively. The coupling
strength is denoted by $\kappa$, the number of oscillators by $N$ and
 {$q$ is a denseness parameter scaling the number of links with
  changing $N$;
\begin{figure}[t]
  \centering
\includegraphics[width=0.99\linewidth]{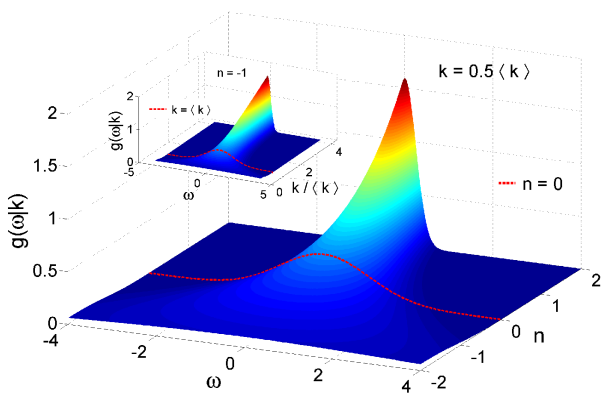}
\caption{(Color online) Conditional Gaussian frequency distribution
  with $\sigma_0=1$ shown as a function of the correlation power $n$ and
  the relative degree $k/\langle k\rangle$ (inset).}
\label{gw}
\end{figure}
for $q=0$ the network is sparse, while it is dense for $q=1$. Such a
normalization is appropriate as long as all the degrees share the same
scaling with the system size. Otherwise one may choose the maximum
degree occurring in the network to guarantee an intensive coupling
term \cite{ArDiazKuMoZh08}. } We consider undirected and
unweighted networks, in which case the adjacency matrix is symmetric
with elements $A_{ij}=1$, if the units $i$ and $j$ are coupled,
otherwise $A_{ij}=0$.  Complex topologies of real-world networks
can be encoded into the adjacency
matrix, and decoded by counting all the degrees, which are given by
\begin{equation}
k_i=\sum_{j=1}^{N} A_{ij},
\label{degrees}
\end{equation}
$k_i>0$ by definition. Calculating the probabilities of occurring degrees, yields the degree distribution $P(k)$. 
 {Various stochastic processes are brought together in the noise terms $\xi_i(t)$, 
such as the variability in the release of neurotransmitters or the quasi-random synaptic inputs from other
neurons \cite{LiGarNeiSchi04}. The sum of stochastic influences is modeled by Gaussian white noise:}
\begin{equation}
\begin{aligned}
\langle\xi_i(t)\rangle&=0,\\
\langle\xi_i(t)\xi_j(t')\rangle&=2D\delta_{ij}\delta(t-t')\,.
\end{aligned}
\end{equation}
The single parameter $D$ scales the noise intensity and is
nonnegative. The angular brackets denote an average over different
realizations of the noise.

\begin{figure*}[ht]
\centering
  \includegraphics[width=0.99\linewidth]{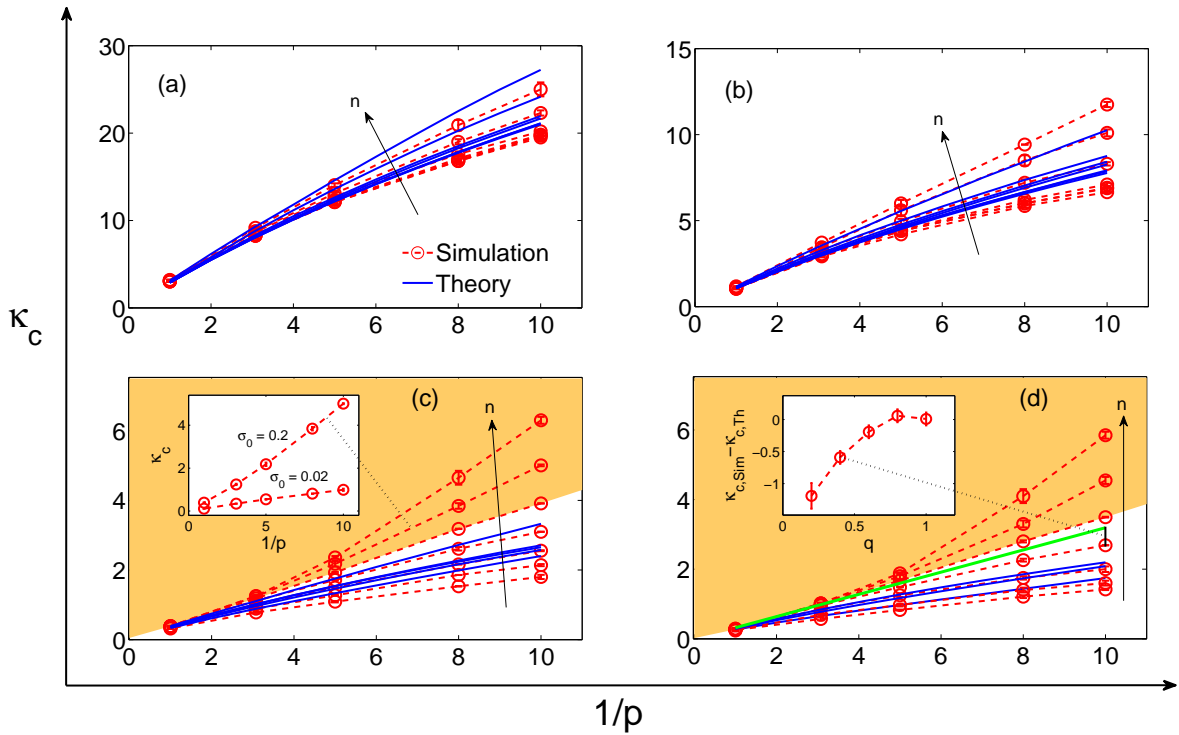}
  \caption[]{(Color online) The critical coupling strength depicted
    as a function of the
    inverse edge probability, which is proportional to the inverse
    average degree, see Eq. \eqref{avg_k}.  Measurements are done
      via finite-size scaling analysis with systems of size
      $N=300,500,800,1200$ and $q=0.4$. Furthermore, a Gaussian
      frequency distribution is used with zero mean and standard
      deviation $\sigma_n(k_i)$, where $\sigma_0=0.2$
      (cf. Eq. \eqref{stdev_b}).  Markers connected by dashed lines
    (red) show simulation results, while (blue) solid lines depict
    corresponding theoretical results, obtained by numerical
    integration. The thick solid (green) line in panel (d) is exactly
    obtained by Eq. \eqref{relations} with $n=1,2$. The arrows show
    the direction of increasing $n$, from $n=-2$ to $n=4$. Panels
    (a)-(d) show results with decreasing noise intensity: (a) $D=1.5$,
    (b) $D=0.5$, (c) $D=0.05$, and (d) $D=0$. Inset in (c) presents
    $\kappa_c$ comparing $\sigma_0=0.2$ with $\sigma_0=0.02$ in case of
    $n=3$. Inset in (d) depicts the discrepancy
    $\kappa_{c,\mathrm{Sim}}-\kappa_{c,\mathrm{Th}}$ between
    simulation and theory as a function of the denseness parameter $q$
    in case of $n=1$ and $p=0.1$. In the unshaded regions of (a)-(d)
    simulation results are accompanied by theoretical results, whereas
    in the shaded areas (orange) in (c) and (d) a superlinear growth
    of $\kappa_c$ cannot be described by our mean-field theory.}
  \label{kq}
\end{figure*}

 There are various possibilities how the individual oscillation
  frequencies and degrees can be correlated, including
  correlations between the mean values or the widths of the
  corresponding distributions. In \cite{Br08,FaHi09} the frequencies and
degrees are assumed to be positively correlated, hence mean values and widths
of the frequency and degree distributions are directly correlated.  It
is found that with increasing positive correlation, the oscillators
are easier to synchronize. This phenomenon gives rise to the fact that
one can observe an abrupt synchronization transition, if the natural
frequencies and degrees are identified \cite{GogaGoArMo11}.

 However, we remark that with respect to many real-world systems, one cannot observe a direct
correlation between individual dynamics and connectivity. In neuronal networks for instance, a higher number of
connections is not directly linked to a higher neuronal firing rate with respect to one cell type. This is due to the balance of inhibition and excitation
\cite{VoSprZeClGer11}, where de- and acceleration compensate each other. Therefore, the central idea in this work is to consider correlations that are due to or affect only the
variability in the degree or the frequency distribution, respectively. Hence, the mean values are not affected. To this end, we assume
\begin{equation}
\sqrt{\langle\omega_i^2\rangle-\langle\omega_i\rangle^2}=\sigma_n(k_i). 
\label{stdev}
\end{equation}
Each oscillator draws its natural frequency from the same distribution function, but with an individual standard deviation, given
by the degree $k_i$. Here, $\sigma_n(k_i)$ is an abbreviation for the power-law function
\begin{equation}
\sigma_n(k_i)=\sigma_0\left(\frac{k_i}{\langle k \rangle}\right)^n, n\in \mathbb R.
\label{stdev_b}
\end{equation}
We call $n$ the correlation power; $\sigma_0$ stands for the original standard deviation without correlations. 
Eq. \eqref{stdev_b} gives rise to two different settings, namely either positive, $n>0$, or negative correlations, $n<0$.
Note that for $k<\langle k \rangle$, i.e. nodes with degrees smaller than the average degree, the natural frequencies are the more
sharply distributed around the mean frequency, the larger is $n$, while for $k>\langle k \rangle$ it is just the opposite case
(see Fig. \ref{gw} for visualization).

We consider Erd\"os-R\'enyi like random networks that are constructed by assigning an edge probability 
\begin{equation}
p_e=p\cdot N^{q-1},\ 0\leq p,q\leq 1
\label{pe}
\end{equation}
for any two of the $N$ nodes in the network with the scaling parameter
$q$ introduced in Eq. \eqref{ourmodel}.  The further additional
requirement is that besides the edge probability, each node is
  a priori connected to another randomly chosen one. In this
way we guarantee that there are no isolated nodes, which are not
interesting here, because they are not able to take part in the
synchronization process and they only reduce the effective system
size.  Hence, the average degree reads
\begin{equation}
\langle k \rangle=2+p N^{q}\left(1-\frac{3}{N}\right),
\label{avg_k}
\end{equation}
which is approximately $p N^{q}$ for $q>0$ and $N\rightarrow\infty$. The first term in Eq. \eqref{avg_k} stems from
the random connections that are a priori chosen, whereas the second term is a result of the edge probabilities,
Eq. \eqref{pe}. Higher moments and the degree distribution $P(k)$ are
not known exactly. However, for large systems and $q>0$, the second term in Eq. \eqref{avg_k} dominates and the degrees become 
binomially distributed \cite{New03}.

In our simulations, the stochastic differential equations are
integrated up to $t=600$ with time step $\Delta t=0.05$ by using the Heun
scheme. We consider a Gaussian frequency distribution
with zero mean and  {standard deviation $\sigma_n(k_i)$, where
  $\sigma_0=0.2$ (cf. Eq. \eqref{stdev_b}).}  Moreover, we discard the data up to $t=200$,
by which transient effects are safely avoided. The statistical
equilibria are further calculated as averages over at least $100$
different network realizations. The different network configurations
do not differ only in the configuration of the connections, but the
oscillators on the network differ as well: all the natural frequencies
and the initial values of the phases change from one configuration to
another.

 In order to measure the critical coupling strength, we perform a
  finite-size scaling analysis \cite{SonnSchi12,HCK02}, where we take
  networks of size $N=300,500,800,1200$ with $q=0.4$. Fig. \ref{kq}
displays the measured critical coupling strengths (red circles). As
expected, both a larger noise intensity $D$
and a decrease of the number of connections, here parameterized by the
inverse edge probability $1/p$, impedes the synchronizability. This
causes the higher coupling strength $\kappa_c$ needed for the onset of
synchronization (compare panels (a)-(d)).

Besides that, we observe that $\kappa_c$ increases
with the correlation power $n$, which is by far not a foregone
conclusion. One could have expected that both settings of correlation,
i.e. $n<0$ and $n>0$ in \eqref{stdev_b}, lead to a decrease of $\kappa_c$,
because the latter marks the transition
from the completely asynchronous to a partially synchronous state, and
not to the completely synchronous state. For any correlation power,
oscillators with a narrower frequency distribution appear which are
easier to synchronize. Hence, a lower coupling strength would be
needed for the onset of synchronization.

Instead, the uncorrelated case $n=0$ needs a critical coupling
strength intermediate to the two settings with $n \ne 0$. Positive correlations
require higher critical coupling strengths
$\kappa_c$, negatively correlated networks can be easier
synchronized, i.e. $\kappa_c$ decays.

First, we provide an intuitive explanation for this observation above; in
the next section a mathematical reasoning will be given. The
phenomenon of synchronization arises by virtue of
interactions. Therefore, nodes with a larger degree $k > \langle k\rangle$ (hubs)
are more crucial than nodes with a smaller degree $k
< \langle k\rangle$ (compare with Ref. \cite{Per10}). 
If the frequencies of these hubs are much broader spread
around the average frequency, it is more difficult for the whole
network to exhibit a synchronized oscillation. In other words, a
population of oscillators is easier to synchronize, if the important
nodes possess frequencies closer to the average frequency. In
particular, for the case $n>0$, hubs are favored to have a great
variability of frequencies, whereas sparsely linked nodes do
the opposite. Necessary coupling for the onset of synchronization has to
be larger than in the uncorrelated case. Differently for $n<0$, the less
linked nodes own an increased variabilty compared to the uncorrelated
case, but the hubs are now easier to synchronize since their
frequencies are narrower distributed.

We further observe that the curves for different $n$ approach each
other with increasing noise intensity $D$. This is due to the fact
that a strong noise outweighs the diversity of the oscillators given
by the frequency distribution; the effect of correlations is destroyed
for large noise intensities and they become negligible.

Finally, it turns out to be beneficial to plot the critical coupling
strength $\kappa_c$ as a function of the inverse edge probability
$1/p$ as done in Fig. \ref{kq}. In this way we observe two distinct
regions: for small correlation power $n$, the critical coupling
strength increases sublinearly as a function of $1/p$, while for large
$n$, it increases superlinearly. In panel (d) a linear dependence is
located between $n=1$ and $n=2$ for $D=0$. For larger noise
intensities or smaller standard deviations $\sigma_0$ (compare inset
in panel (c)), the separation between the two regions
appears for larger correlation powers $n$.  The shaded areas (orange)
in panels (c) and (d) depict up to which $n$ we find
the superlinear region by numerical simulations.

A linear dependence indicates that the onset of
  synchronization $\kappa_c$ solely depends on the mean degree $\langle k\rangle$,
  which is determined by $p$. Hence, for a certain $n_c$, the
  onset of synchronization seems to become independent of higher moments of
  the degree distribution. The heterogeneity in the network is masked by the
  correlations.

\section{Theoretical considerations}
\label{analyt}
In what follows, we discuss an approximation scheme \cite{SonnSchi12}
  that allows to reproduce analytically our observations a\-bove. We
replace the random network by a fully connected network with random
coupling weights that mimic the actual network structure.  Requiring
thereby the conservation of the individual degrees,
$k_i=\sum_{j=1}^{N} \tilde A_{ij},\ i=1,\ldots,N$
(cf. Eq. \eqref{degrees}), the elements of the approximated adjacency
matrix read
\begin{equation}
  \tilde A_{ij}\,=\, k_i \, \frac{k_j}{\sum_{l=1}^{N} k_{l}}.
\label{newA}
\end{equation}
Inserting this into Eq. \eqref{ourmodel} yields a weighted mean-field
approximation and effectively a one-oscillator description
\cite{SonnSchi12}. In the following we consider the thermodynamic
limit $N\rightarrow\infty$, where the system is conveniently described
by the density $\rho(\phi,t|\omega,k)$, which is normalized according
to $\int_0^{2\pi}\rho(\phi,t|\omega,k)\mathrm d\phi=1\ \forall\
\omega,k,t$.

For given degree $k$ and natural frequency $\omega$,
$\rho(\phi,t|\omega,k)$ $\mathrm d\phi$ gives the fraction of oscillators
having a phase between $\phi$ and $\phi+\mathrm d\phi$ at time $t$
(indices can be neglected, since all the nodes are assumed to be
statistically identical).
The completely asynchronous state is given by
$\rho(\phi,t|\omega,k)=1/(2\pi)\ \forall\ \omega,k,t$ and we aim at
calculating the critical coupling strength, where it loses its
stability, which marks the onset of synchronization. The linear
stability of the completely asynchronous state is characterized by a
single real-valued eigenvalue $\lambda$ given by a self-consistent equation
\cite{SonnSchi12}:
\begin{equation}
  1=\frac{\kappa}{2 N^q\langle k\rangle}
  \int_{-\infty}^{+\infty}\mathrm d\omega'\sum_{k'}\frac{(\lambda+D) k'^2}
  {(\lambda+D)^2+\omega'^2}P\left(\omega',k'\right)\ .
  \label{selfconsist}
\end{equation}
The sum over $k'$ covers all possible degrees, which could be further
approximated by an integral.  The joint probability density
$P(\omega,k)$ takes into account the possibility of correlations
between the frequencies and degrees.  In the derivation of
Eq. \eqref{selfconsist} we assume that, with regard to the
$\omega$-dependency, $P(\omega,k)$ has a single maximum at frequency
$\omega=0$ (this choice is always possible due to the rotational symmetry)
and is symmetric with respect to it.

The critical condition $\lambda=\lambda_c=0$ yields the critical
coupling strength
\begin{equation}
  \kappa_c=2 N^q\langle k\rangle\left[\int_{-\infty}^{+\infty}\mathrm d\omega'\sum_{k'}\frac{D k'^{2}}{D^2+\omega'^2}P\left(\omega',k'\right)\right]^{-1}\ .
\label{kappac}
\end{equation}
This equation is not valid in the noise-free case, where one has to
take the limit $\lambda\to 0^{+}$ in Eq. \eqref{selfconsist} with
$D=0$ resulting in
\begin{equation}
  \kappa_c=2 N^q\langle k\rangle\left[\pi\sum_{k'}k'^{2}P\left(0,k'\right)\right]^{-1}\ .
\label{kappac_D0}
\end{equation}
To see this, note that $\lim_{\lambda \to
  0^{+}}\int_{-\infty}^{+\infty}\mathrm
d\omega'\lambda/\left(\lambda^2+\omega'^2\right)=\pi\int_{-\infty}^{+\infty}\mathrm
d\omega'\delta(\omega')$ \cite{StrMir91}.

Assuming a given degree distribution $P(k)$, the joint frequency and
degree distribution separates as $P(\omega,k)\equiv g(\omega|k)P(k)$
with the conditional frequency distribution $g(\omega|k)$. 
It gives the probability that an oscillator at a node with
degree $k$ has the natural frequency $\omega$. It includes the
relation \eqref{stdev}, i.e. the correlations between the degree and
the frequency variation.

First, in accordance with the numerics above, we consider a Gaussian
frequency distribution:
\begin{equation}
  g_{\rm{gauss}}(\omega|k)=\frac{1}{\sqrt{2\pi}\sigma_n(k)}\mathrm e^{-\frac{1}{2}\frac{\omega^2}{\sigma_n(k)^2}}.
\label{gwk}
\end{equation}
and $\sigma_n(k)$ is expressed by \eqref{stdev}. Taking the integral,
we derive the critical coupling strength \eqref{kappac}:
\begin{equation}
\begin{aligned}
\kappa_{c,\rm{gauss}}=&\ 2\sqrt{\frac{2}{\pi}}\sigma_0 N^q\langle k\rangle^{1-n}\\
&\times\left\langle k^{2-n}\operatorname{erfc}\left(\frac{D}{\sqrt{2}\sigma_n(k)}\right)\exp\left(\frac{D^2}{2\sigma_n(k)^2}\right)\right\rangle^{-1},
\end{aligned}
\label{kcgaussD}
\end{equation}
which is an intensive parameter and scales with the variation of
  frequencies for small $\sigma_0$ (see inset in panel (c) of
  Fig. \ref{kq}).

By calculating $\mathrm d\kappa_c/\mathrm d n$, we want to validate that the critical coupling strength indeed grows with the correlation power,
as stated in the previous section. To this end, we use Eq. \eqref{kcgaussD} and arrive at a sufficient condition for $\mathrm d\kappa_{c,\rm{gauss}}/\mathrm d n>0$, namely
\begin{equation}
\frac{2}{\sqrt{\pi}}y>\left(2y^2-1\right)\exp\left(y^2\right)\operatorname{erfc}\left(y\right),
\label{ineq}
\end{equation}
with $y=D\left(\langle k\rangle/k\right)^n/\left(\sqrt{2}\sigma_0\right)$. Since we have $y>0$, the inequality \eqref{ineq} is true; in fact,
the right-hand side divided by $y$, approaches $2/\sqrt{\pi}$ from below for $y$ going to infinity.

In the noise-free case $D=0$ we find
\begin{equation}
\kappa_{c,\rm{gauss}}(D=0)=2\sqrt{\frac{2}{\pi}}\sigma_0 N^q\frac{\langle k\rangle^{1-n}}{\langle k^{2-n}\rangle}.
\label{relations}
\end{equation}
\\*Interestingly, for $n=1$ and $n=2$, we get the same critical coupling
strength growing inversely to the average degree:
\begin{equation}
\kappa_{c,\rm{gauss}}(D=0,n=1,2)=2\sqrt{\frac{2}{\pi}}\frac{\sigma_0}{p}
\label{kcgaussD0}
\end{equation}
in the thermodynamic limit (cf. Eq. \eqref{avg_k}). In Fig. \ref{kq}
the solid blue lines
  describe the critical coupling strength as given by numerical
  integration of Eqs. \eqref{kcgaussD} or \eqref{relations} with a binomial degree distribution
  and system size $N=1000$. 
In panel (d) for $n=1,2$, instead of the numerical integration of Eq. \eqref{relations},
the exact expression \eqref{kcgaussD0} is shown (thick green line).
The agreement between theory and simulation is
  satisfactory in (a)-(d) and confirms the previous
  observations. The weighted mean-field
  approximation does not yield superlinear dependencies. Simulation results in the shaded
  (orange) areas in panels (c) and (d) are therefore not covered by the theory; the
  validity of the approximation restricts to correlation strengths
  with sub- and linear growth of $\kappa_c$.
  For large noise intensities $D$ the theory overestimates the critical coupling strength $\kappa_c$, irrespective
  of the correlation power $n$, but this may turn into the opposite case when decreasing $D$ depending on $n$.
  In summary, there seems to be some particular noise values where the
agreement between theory and simulation is particularly good.
As presented in the inset of panel (d), deviations between the results
from the numerical simulations $\kappa_{c,\mathrm{Sim}}$ and from the weighted
mean-field theory $\kappa_{c,\mathrm{Th}}$ can be further reduced by increasing
$q$. Hence, more densely connected networks are better reflected by
the theory.

\section{Generalizations}
The network model under consideration constitutes already a
generalized model, since it allows to interpolate between sparse and
dense random networks.  Here we discuss two further
generalizations, namely other frequency distributions and different
normalization variants of the coupling term,
cf. Eq. \eqref{ourmodel}. In particular, we consider now a Lorentzian
and a uniform frequency distribution:
\begin{equation}
\begin{aligned}
g_{\rm{lorentz}}(\omega|k)&=\frac{\sigma_n(k)}{\pi}\frac{1}{\sigma_n(k)^2+\omega^2}, \\
g_{\rm{uni}}(\omega|k)&=\frac{1}{2\sqrt{3}\sigma_n(k)},\ |\omega|\leq\sqrt{3}\sigma_n(k),
\label{gwk_b}
\end{aligned}
\end{equation}
Note that in case of the Lorentzian, $\sigma_n(k)$ does not have the meaning of a standard deviation, instead it is the scale parameter for the
width of the distribution. We further introduce a generalized normalization $\mathcal N(k)$ instead of $N^q$, which can be a function of the degree $k$.
Then we find for the critical coupling strength:
\begin{equation}
\begin{aligned}
&\kappa_{c,\rm{lorentz}}=2\langle k\rangle\left\langle\frac{k^{2}}{\mathcal N(k)}\frac{1}{D+\sigma_n(k)}\right\rangle^{-1},\\
&\kappa_{c,\rm{uni}}=2\sqrt{3}\sigma_0\langle k\rangle^{1-n}\left\langle\frac{k^{2-n}}{\mathcal N(k)}\arctan\left(\frac{\sqrt{3}\sigma_n(k)}{D}\right)\right\rangle^{-1}.
\end{aligned}
\end{equation}
Let us now specify the normalization $\mathcal N(k)$ by considering two cases:
$\mathcal N(k)=\langle k\rangle$ and $\mathcal N(k)=k$. In the first case, one assumes again that the system-size scaling of the number
of connections is the same for all nodes. In order to distinguish the two normalizations,
we denote the critical coupling strength by $\kappa_a$ or $\kappa_w$, respectively. In the noise-free case $D=0$ we obtain, in contrast to \eqref{relations},
the following relations with the same constant of proportionality $C$:

\begin{equation}
\kappa_a=C\sigma_0\frac{\langle k\rangle^{2-n}}{\langle k^{2-n}\rangle},\ 
\kappa_w=C\sigma_0\frac{\langle k\rangle^{1-n}}{\langle k^{1-n}\rangle},
\label{relations_b}
\end{equation}
irrespective of whether we consider a Gaussian, a Lorent\-zian or a uniform frequency distribution. Only the constant of proportionality $C$ is different,
namely $C_{\rm{gauss}}=2\sqrt{2/\pi}\approx 1.60$, $C_{\rm{lorentz}}=2$ or $C_{\rm{uni}}=4\sqrt{3}/\pi\approx 2.21$.

Again we find a disappearance of network effects for specific correlation powers. The phenomenon is even more pronounced here, since $\kappa_a$ becomes
a constant for $n=1,2$. Moreover we see that $\kappa_w(n)=\kappa_a(n+1)$. Correspondingly, $\kappa_w$ becomes a constant for $n=0,1$. It has been pointed out
in the literature, e.g. in \cite{ArDiazKuMoZh08}, that the additional weight introduced by the normalization $\mathcal N(k)=k$ can mask the heterogeneity of the network. Our
results constitute a generalization of this statement. Preliminary numerical simulations can reproduce our theoretical result $\kappa_w(n)=\kappa_a(n+1)$, while the point
where the onset of synchronization loses its dependence on the network connectivity is found to appear at smaller values of $n$ than predicted by the theory.

\section{Conclusion}
We assumed correlations between the degree number in a complex betwork
and the variance of frequencies of phase oscillators belonging to the
nodes. By estimating numerically and analytically the critical
coupling strength that marks the onset of synchronization, we were
able to show that correlations can favor as well as impede the ability
of creating coherent network oscillations. In both scenarios, the
behavior of hubs (nodes with large degree) plays a dominant
role. Stronger coupling is necessary, if the hubs have broadly
distributed frequencies. The onset of synchonization is shifted up,
despite the fact that the less linked nodes possess a narrower
frequency band and would, taken separately, synchronize at lower
couplings. The opposite happens in case that hubs have narrowed distributions.
We have further demonstrated that noise acting on the frequencies
plays a crucial role; correlation effects become maximally strong in
case of vanishing noise intensity.

We mention that our analysis was performed for edge probabilities $p$
larger than $0.1$. In this region we have found a good applicability
of the weighted mean field theory proposed in
\cite{SonnSchi12}. Analytical results agree satisfactorily with
numeric ones for sufficiently large noise $D$ or not too strong
correlation powers $n$ and dense enough scaling of the links $q>0$ in
particular, as long as the critical coupling grows sub- or linearly
with $1/p$. Beyond a certain denseness parameter for given correlation
power, the weighted mean-field approximation breaks down.

In a recent preprint \cite{SkSuTaRes12},
the masking of the structural heterogeneity was independently found
for another type of correlation.

\begin{acknowledgement}
BS gratefully acknowledges support from the
GRK1589/1, P.K. Radtke for a critical reading of the manuscript and I. Segev 
for fruitful discussions. LSG acknowledges support
by the Bernstein Center Berlin (Project No. A3). FS acknowledges M.A. Serrano 
for fruitful discussions.
\end{acknowledgement}

 \bibliography{bibliography}
\bibliographystyle{unsrt}

\end{document}